\documentclass[twocolumn,times]{aastex6}
\usepackage{amsmath}
\usepackage{multirow}
\usepackage{textcomp}
\usepackage[nolist]{acronym}

\usepackage{listings}
\DeclareMathOperator{\erf}{erf}
\DeclareMathOperator{\erfc}{erfc}

\newcommand\transpose{\ensuremath{^{^\mathsf{T}}}}

\begin{document}

\title{Supplement: ``Going the Distance: Mapping Host Galaxies of LIGO and Virgo Sources in Three Dimensions Using Local Cosmography and Targeted Follow-up''}
\shorttitle{Supplement: Advanced LIGO/Virgo Volume Reconstruction and Galaxy Catalogs}

\slugcomment{The Astrophysical Journal Supplement Series, 226:10, 2016 September}
\received{2016 May 13}
\accepted{2016 August 5}
\published{2016 September 21}

\AuthorCallLimit=-1

\author{Leo~P.~Singer\altaffilmark{1,2}}
\author{Hsin-Yu~Chen\altaffilmark{3}} \author{Daniel~E.~Holz\altaffilmark{3}} \author{Will~M.~Farr\altaffilmark{4}} \author{Larry~R.~Price\altaffilmark{5}} \author{Vivien~Raymond\altaffilmark{5,6}} \author{S.~Bradley~Cenko\altaffilmark{1,7}} \author{Neil~Gehrels\altaffilmark{1}} \author{John~Cannizzo\altaffilmark{1}} \author{Mansi~M.~Kasliwal\altaffilmark{8}} \author{Samaya~Nissanke\altaffilmark{9}} \author{Michael~Coughlin\altaffilmark{10}} \author{Ben~Farr\altaffilmark{3}} \author{Alex~L.~Urban\altaffilmark{11}} \author{Salvatore~Vitale\altaffilmark{12}} \author{John~Veitch\altaffilmark{4}} \author{Philip~Graff\altaffilmark{13}}
\author{Christopher~P.~L.~Berry\altaffilmark{4}} \author{Satya~Mohapatra\altaffilmark{12}} \author{Ilya~Mandel\altaffilmark{4}} 
\altaffiltext{1}{Astroparticle Physics Laboratory, NASA Goddard Space Flight Center, Mail Code 661, Greenbelt, MD 20771, USA}
\altaffiltext{2}{NASA Postdoctoral Program Fellow}
\altaffiltext{3}{Department of Physics, Enrico Fermi Institute, and Kavli Institute for Cosmological Physics, University of Chicago, Chicago, IL 60637, USA}
\altaffiltext{4}{School of Physics and Astronomy, University of Birmingham, Birmingham, B15 2TT, UK}
\altaffiltext{5}{LIGO Laboratory, California Institute of Technology, Pasadena, CA 91125, USA}
\altaffiltext{6}{Albert-Einstein-Institut, Max-Planck-Institut f\"ur Gravitationsphysik, D-14476 Potsdam-Golm, Germany}
\altaffiltext{7}{Joint Space-Science Institute, University of Maryland, College Park, MD 20742, USA}
\altaffiltext{8}{Cahill Center for Astrophysics, California Institute of Technology, Pasadena, CA 91125, USA}
\altaffiltext{9}{Institute of Mathematics, Astrophysics and Particle Physics, Radboud University, Heyendaalseweg 135, 6525 AJ Nijmegen, The Netherlands}
\altaffiltext{10}{Department of Physics and Astronomy, Harvard University, Cambridge, MA 02138, USA}
\altaffiltext{11}{Leonard E. Parker Center for Gravitation, Cosmology, and Astrophysics, University of Wisconsin--Milwaukee, Milwaukee, WI 53201, USA}
\altaffiltext{12}{LIGO Laboratory, Massachusetts Institute of Technology, 185 Albany Street, Cambridge, MA 02139, USA}
\altaffiltext{13}{Department of Physics, University of Maryland, College Park, MD 20742, USA}

\shortauthors{Singer et al.}

\keywords{
gravitational waves
---
galaxies: distances and redshifts
---
catalogs
---
surveys
}

\newcommand{\dd}{\ensuremath{\mathrm{d}}}

\defcitealias{GoingTheDistance}{Letter}
\defcitealias{GoingTheDistanceSupplement}{Supplement}
 
\acused{HEALPix}
\acused{2+1D}
\acused{2D}
\acused{3D}
\acused{BAYESTAR}

\begin{abstract}
This is an online supplement to the Letter of \citeauthor{GoingTheDistance}, in which we demonstrated a rapid algorithm for obtaining joint \ac{3D} estimates of sky location and luminosity distance from observations of \acl{BNS} mergers with Advanced \acs{LIGO} and Virgo. We argued that combining the reconstructed volumes with positions and redshifts of possible host galaxies can provide large-aperture but small \acl{FOV} instruments with a manageable list of targets to search for optical or infrared emission. In this Supplement, we document the new \acs{HEALPix}-based file format for \acs{3D} localizations of gravitational-wave transients. We include Python sample code to show the reader how to perform simple manipulations of the \ac{3D} sky maps and extract ranked lists of likely host galaxies. Finally, we include mathematical details of the rapid volume reconstruction algorithm.
\end{abstract}

\section{Outline of This Supplement}

In \citet{GoingTheDistance}, we discussed the measurement of luminosity distances of \ac{CBC} events using the Advanced \ac{LIGO} and Virgo ground-based interferometric gravitational-wave (\acsu{GW}) detectors. In the main Letter, an algorithm was introduced for rapidly extracting directionally dependent distance estimates from \ac{GW} observations and illustrated the typical \ac{3D} shape of \ac{GW} volume reconstructions during early Advanced \ac{LIGO}. Finally, we argued that the \ac{3D} structure and distance information can be leveraged to guide searches of likely nearby host galaxies for X-ray, optical, and infrared counterparts of \ac{BNS} mergers.

This Supplement provides the following supporting material. First, in \S\ref{sec:format}, we document a file format for the rapid transmission of \ac{3D} volume reconstructions in \ac{GW} alerts. It is based on and is backward-compatible with the \ac{2D} localization formation that we introduced in \citet{FirstTwoYears} and that was employed in \ac{GW} alerts that were sent in \acl{O1} (\acsu{O1}; \citealt{GW150914-EMFOLLOW}). Second, in \S\ref{sec:data}, we describe the online data release, which provides a browsable collection of simulated \ac{3D} sky maps. Third, in \S\ref{sec:tutorial}, we provide a Python primer for performing basic operations on \ac{3D} sky maps, all the way through selecting a list of the most likely host galaxies. In \S\ref{sec:algorithm}, we provide additional details of the position reconstruction algorithm. Finally, in \S\ref{sec:faithfulness}, we show that the algorithm produces faithful representations of the full \ac{3D} probability distributions.

The reader who is interested in leveraging \ac{GW} distance information for planning \ac{EM} follow-up observations or performing archival research needs only consult \S\ref{sec:format} and \S\ref{sec:tutorial}.

\section{3D Localization File Format}
\label{sec:format}

The \ac{3D} localization for a single \ac{GW} candidate is stored as a \acl{FITS} (\acsu{FITS}; \citealt{FITS}) file. The \ac{FITS} file contains a single binary table \citep{FITSBinTable} that represents a \acl{HEALPix} (\acsu{HEALPix}; \citealt{healpix}) all-sky image. The table has four floating-point columns, listed in Table~\ref{table:fitscolumns}, which represent four channels of the \ac{HEALPix} image. The first column, \texttt{PROB}, is simply the probability that the source is contained within the pixel $i$ that is centered on the direction $\boldsymbol{n}_i$, the same as in the \ac{2D} localization format. The second and third columns, \texttt{DISTMU} and \texttt{DISTSTD}, are the ansatz location and scale parameters, respectively. The fourth column, \texttt{DISTNORM}, is the ansatz normalization coefficient, included for convenience.

In pixels on the sky that contain very little probability, sometimes the conditional distance distribution cannot be represented using the ansatz. This is signaled by \texttt{DISTMU}=$\infty$, \texttt{DISTSIGMA}=$1$, and \texttt{DISTNORM}=0.

\begin{deluxetable*}{>{\ttfamily}llll}
    \tablecolumns{4}
    \tablewidth{\textwidth}
    \tabletypesize{\normalsize}
    \tablecaption{\label{table:fitscolumns}\ac{HEALPix} Columns}
    \tablehead{
        \colhead{FITS Name} &
        \colhead{Symbol} &
        \colhead{Units} &
        \colhead{Description}
    }
    \startdata
        PROB & $\rho_i$ & pixel$^{-1}$ & Probability that the source is contained in pixel $i$, centered on the direction $\boldsymbol{n}_i$ \\
        DISTMU & $\hat\mu_i$ & Mpc & Ansatz location parameter of conditional distance distribution in direction $\boldsymbol{n}_i$, or $\infty$ if invalid \\
        DISTSIGMA & $\hat\sigma_i$ & Mpc & Ansatz scale parameter of conditional distance distribution in direction $\boldsymbol{n}_i$, or $1$ if invalid \\
        DISTNORM & $\hat{N}_i$ & Mpc$^{-2}$ & Ansatz normalization coefficient, or $0$ if invalid
    \enddata
\end{deluxetable*}

The \ac{FITS} header, an example of which is shown in Table~\ref{table:fitsheader}, provides metadata including the UTC time of the \ac{GW} trigger and the list of \ac{GW} instruments that contributed to the localization. The header also provides values for \texttt{DISTMEAN} and \texttt{DISTSTD}, respectively, being the posterior mean and standard deviation of distance marginalized over the whole sky.

\begin{deluxetable*}{>{\ttfamily}l>{\ttfamily}l>{\ttfamily}l}
    \tablecolumns{3}
    \tablewidth{\textwidth}
    \tabletypesize{\footnotesize}
    \tablecaption{\label{table:fitsheader}Example FITS Header}
    \tablehead{
        \colhead{Key} &
        \colhead{Value} &
        \colhead{Comment}
    }
    \startdata
        \cutinhead{HDU 0}
        SIMPLE  &                    T & conforms to FITS standard                      \\
        BITPIX  &                    8 & array data type                                \\
        NAXIS   &                    0 & number of array dimensions                     \\
        EXTEND  &                    T                                                  \\
        \cutinhead{HDU 1}
        XTENSION& 'BINTABLE'           & binary table extension                         \\
        BITPIX  &                    8 & array data type                                \\
        NAXIS   &                    2 & number of array dimensions                     \\
        NAXIS1  &                16384 & length of dimension 1                          \\
        NAXIS2  &                 3072 & length of dimension 2                          \\
        PCOUNT  &                    0 & number of group parameters                     \\
        GCOUNT  &                    1 & number of groups                               \\
        TFIELDS &                    4 & number of table fields                         \\
        TTYPE1  & 'PROB    '                                                            \\
        TFORM1  & '1024E   '                                                            \\
        TUNIT1  & 'pix-1   '                                                            \\
        TTYPE2  & 'DISTMU  '                                                            \\
        TFORM2  & '1024E   '                                                            \\
        TUNIT2  & 'Mpc     '                                                            \\
        TTYPE3  & 'DISTSIGMA'                                                           \\
        TFORM3  & '1024E   '                                                            \\
        TUNIT3  & 'Mpc     '                                                            \\
        TTYPE4  & 'DISTNORM'                                                            \\
        TFORM4  & '1024E   '                                                            \\
        TUNIT4  & 'Mpc-2   '                                                            \\
        PIXTYPE & 'HEALPIX '           & HEALPIX pixelisation                           \\
        ORDERING& 'NESTED  '           & Pixel ordering scheme, either RING or NESTED   \\
        COORDSYS& 'C       '           & Ecliptic, Galactic or Celestial (equatorial)   \\
        EXTNAME & 'xtension'           & name of this binary table extension            \\
        NSIDE   &                  512 & Resolution parameter of HEALPIX                \\
        FIRSTPIX&                    0 & First pixel \# (0 based)                       \\
        LASTPIX &              3145727 & Last pixel \# (0 based)                        \\
        INDXSCHM& 'IMPLICIT'           & Indexing: IMPLICIT or EXPLICIT                 \\
        OBJECT  & 'coinc\_event:coinc\_event\_id:18951' & Unique identifier for this event \\
        INSTRUME& 'H1,L1   '           & Instruments that triggered this event          \\
        DATE-OBS& '2010-09-03T06:12:26.60324' & UTC date of the observation             \\
        MJD-OBS &     55442.2586412414 & modified Julian date of the observation        \\
        DATE    & '2015-04-13T10:17:11' & UTC date of file creation                     \\
        CREATOR & 'bayestar\_localize\_coincs.py' & Program that created this file      \\
        DISTMEAN&   68.54061620909769  & Posterior mean distance in Mpc \\
        DISTSTD &   17.14006463067744  & Posterior standard deviation of distance in Mpc \\
    \enddata
\end{deluxetable*}

\section{Data Release}
\label{sec:data}

\begin{figure*}
    \includegraphics[width=\textwidth]{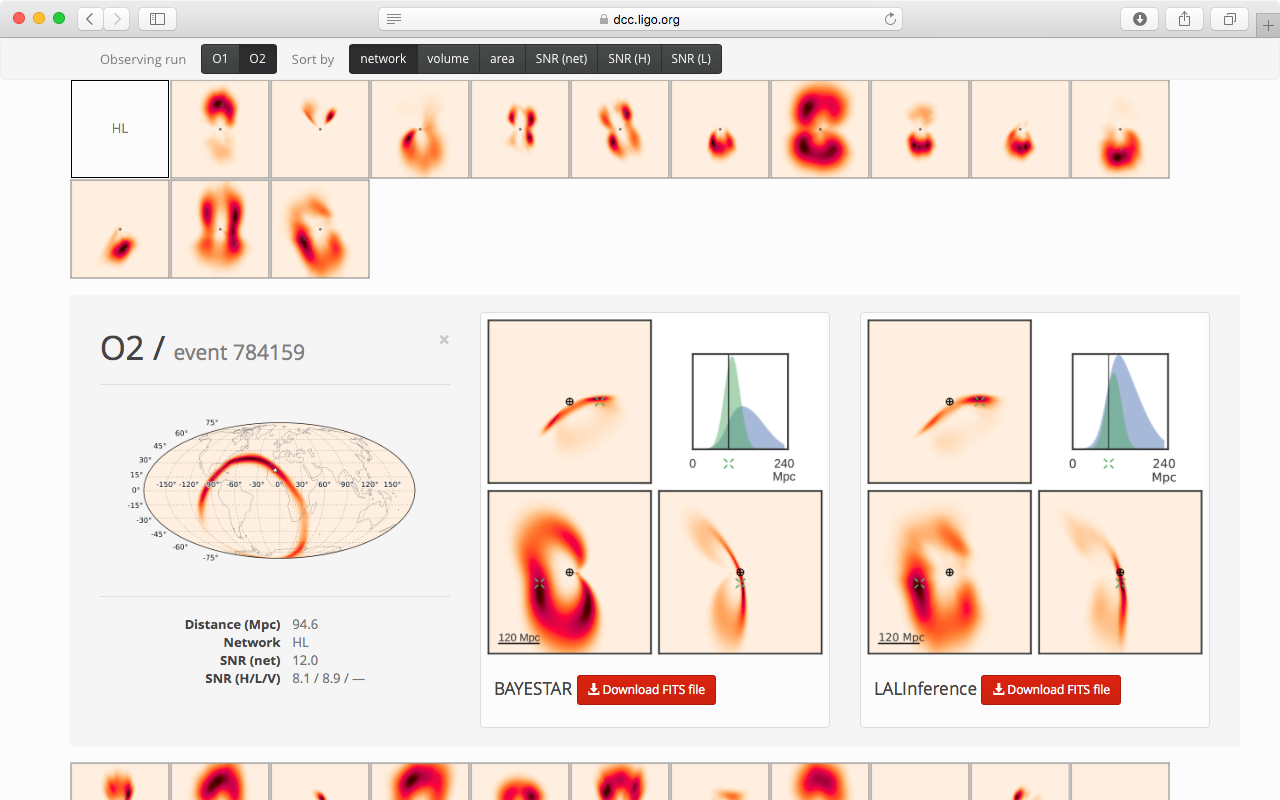}
    \caption{\label{fig:screenshot}Screen shot of data release page.}
\end{figure*}

An online data release provides a browsable catalog of simulated \ac{3D} \ac{GW} localizations. One may select events from \ac{O1} or \ac{O2}. Events may be sorted by detector network (a one- or two-letter combination consisting of `H' for \acl{LHO}, `L' for \acl{LLO}, `V' for Virgo), 90\% credible volume in Mpc$^3$, 90\% credible area in deg$^2$, or \ac{SN}. For each event, a \ac{BAYESTAR} or LALInference \ac{FITS} file may be downloaded. A screen shot of the data release is shown in Fig.~\ref{fig:screenshot}.

\section{Python Example Code}
\label{sec:tutorial}

In this section, we provide some Python sample code to perform some simple manipulations of \ac{3D} sky maps. The triple greater-than signs (\texttt{>>>}~) and triple-dots (\texttt{...}~) are the Python interactive prompt; the reader should type everything on the line \emph{after} these.

\subsection{Python Environment}

These examples will work in Python 2.7 and later on Linux or UNIX systems. If the reader does not already have a Python environment of preference, we suggest the Anaconda Python distribution\footnote{\url{https://www.continuum.io/anaconda}} for desktop use or the lightweight Miniconda variant\footnote{\url{http://conda.pydata.org/miniconda.html}} for computing clusters. Only the Astropy, Healpy, and Numpy packages are essential for working with the \ac{3D} localizations, but the examples below will also use Matplotlib, Scipy, and Astroquery. All of these packages can be installed with Pip\footnote{\url{https://pip.pypa.io}}:
\begin{lstlisting}
$ pip install astropy astroquery healpy matplotlib scipy
\end{lstlisting}

\subsection{Reading Sky Maps}

For all of the samples below, start by importing the Healpy for working with \ac{HEALPix} files, the Numpy for vector operations, Matplotlib for plotting, and Scipy for probability functions:
\begin{lstlisting}
$ python
>>> import healpy as hp
>>> import numpy as np
>>> from matplotlib import pyplot as plt
>>> from scipy.stats import norm
\end{lstlisting}

Next, select download an example sky map from the data release. In this example, we use the simulated event that is shown in Figs.~1 and 2 of \citet{GoingTheDistance}. A convenient way to download it is using Astropy's \texttt{download\_file} utility, which will retrieve the file and cache it locally:
\begin{lstlisting}
>>> from astropy.utils.data import download_file
>>> url = ('https://dcc.ligo.org/P1500071/public'
... + '/18951_bayestar.fits.gz')
>>> filename = download_file(url, cache=True)
\end{lstlisting}

The new \ac{3D} localization format is backward-compatible with the \ac{2D} format introduced in \citet{FirstTwoYears}. By default, when we read the HEALPix file with Healpy (or any other common-place HEALPix library or tool), we get just the first layer, the probability sky map:
\begin{lstlisting}
>>> prob = hp.read_map(filename)
\end{lstlisting}

To read both the probability layer and the three additional distance layers, we need to pass the optional \texttt{field=} parameter to Healpy:
\begin{lstlisting}
>>> prob, distmu, distsigma, distnorm = hp.read_map(
... filename, field=[0, 1, 2, 3])
\end{lstlisting}
or slightly more concisely:
\begin{lstlisting}
>>> prob, distmu, distsigma, distnorm = hp.read_map(
... filename, field=range(4))
\end{lstlisting}

Last, it will be useful for subsequent Healpy calls to have the \ac{HEALPix} resolution on hand:
\begin{lstlisting}
>>> npix = len(prob)
>>> npix
3145728
>>> nside = hp.npix2nside(npix)
>>> nside
512
\end{lstlisting}

\subsection{\ac{2D} Probability in a Given Line of Sight}

In this example, we compute the \ac{2D} probability per steradian or per deg$^2$ that the source is in a given direction. Let's take as an example the following equatorial coordinates:
\begin{lstlisting}
>>> ra, dec = 137.8, -39.9
\end{lstlisting}
which, coincidentally, happen to be the true simulated position to the source.

Healpy uses ``physicist's" spherical coordinates ($\theta, \phi$), with $\theta \in [0, \pi]$ being the colatitude from the north celestial pole in radians, and $\phi \in [0, 2\pi)$ being the right ascension in radians. We convert
\begin{lstlisting}
>>> theta = 0.5 * np.pi - np.deg2rad(dec)
>>> phi = np.deg2rad(ra)
\end{lstlisting}

Next, we use Healpy to look up the index of the \ac{HEALPix} pixel that contains that direction:
\begin{lstlisting}
>>> ipix = hp.ang2pix(nside, theta, phi)
>>> ipix
2582288
\end{lstlisting}

Healpy will tell us the area per pixel in steradians at the current \ac{HEALPix} resolution:
\begin{lstlisting}
>>> pixarea = hp.nside2pixarea(nside)
>>> pixarea
3.994741635118857e-06
\end{lstlisting}
or in deg$^2$:
\begin{lstlisting}
>>> pixarea_deg2 = hp.nside2pixarea(nside, degrees=True)
>>> pixarea_deg2
0.013113963206424481
\end{lstlisting}

All that is left to do is look up the probability contained within pixel \texttt{ipix} and (if desired) divide by the area per pixel to obtain the probability per steradian:
\begin{lstlisting}
>>> dp_dA = prob[ipix] / pixarea
>>> dp_dA
7.4387317043042076
\end{lstlisting}
or the probability per deg$^2$:
\begin{lstlisting}
>>> dp_dA_deg2 = prob[ipix] / pixarea_deg2
>>> dp_dA_deg2
0.0022659672582507331
\end{lstlisting}

\subsection{Conditional Distance Distribution Along a Line of Sight}

Next, we calculate the conditional distance distribution along a given line of sight, which is the probability per unit distance under the assumption that the source is in a given direction. We will use the same sky position as in the example above. We lay out a grid in distance along that line of sight:
\begin{lstlisting}
>>> r = np.linspace(0, 150)
\end{lstlisting}

Then, we plug everything into the ansatz distribution:
\begin{lstlisting}
>>> dp_dr = r**2 * distnorm[ipix] * norm(
... distmu[ipix], distsigma[ipix]).pdf(r)
\end{lstlisting}

Finally, we plot the result:
\begin{lstlisting}
>>> plt.plot(r, dp_dr)
>>> plt.xlabel('distance (Mpc)')
>>> plt.ylabel('prob Mpc$^{-1}$')
>>> plt.show()
\end{lstlisting}
\includegraphics[width=\columnwidth]{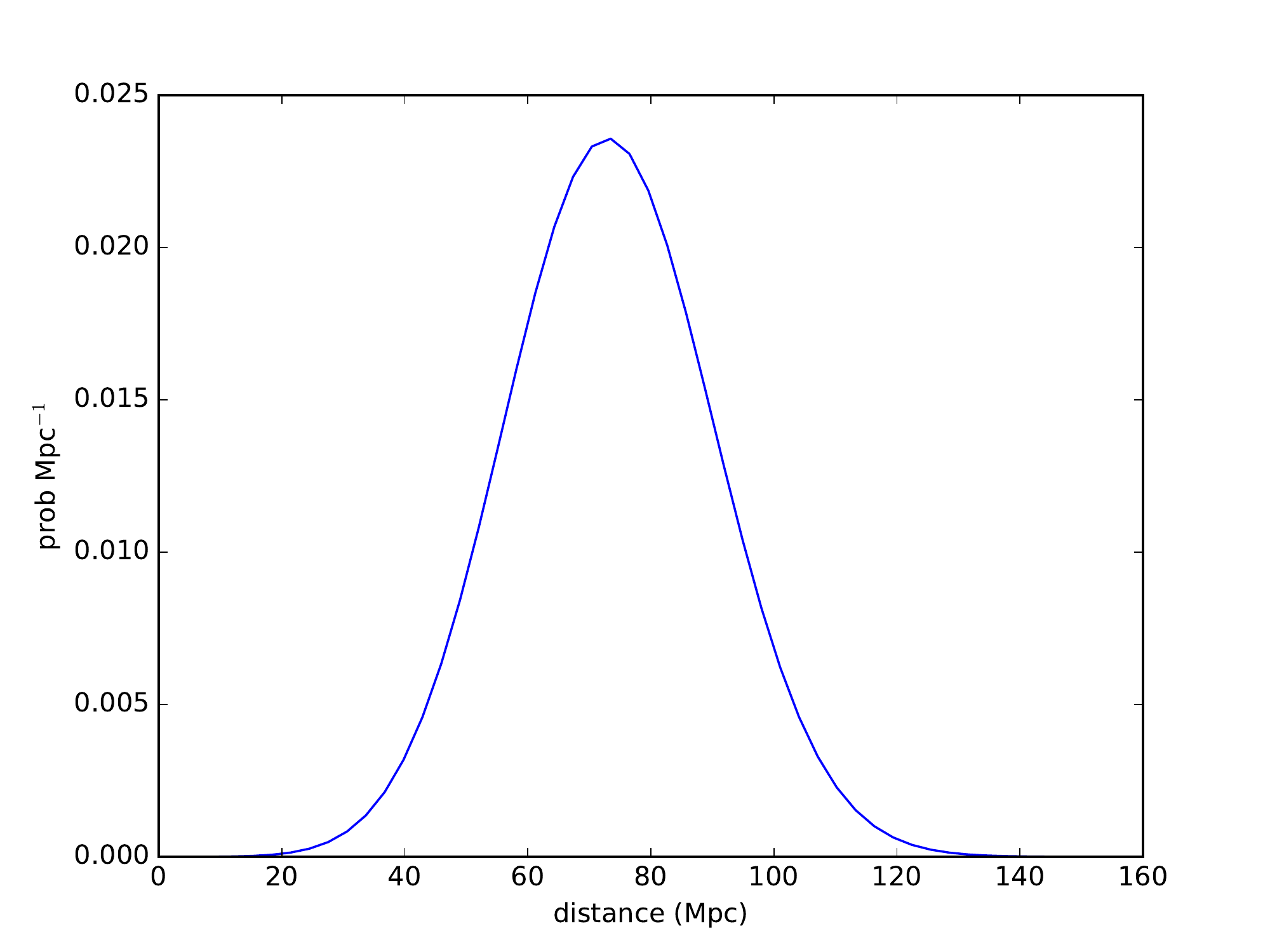}

\subsection{Probability per Unit Volume at a Point}

Now, we calculate the probability density per Mpc$^3$ at a point. We will use the same right ascension and declination as above and a distance of 74.8\,Mpc:
\begin{lstlisting}
>>> r = 74.8
\end{lstlisting}

Finally,
\begin{lstlisting}
>>> dp_dV = prob[ipix] * distnorm[ipix] * norm(
... distmu[ipix], distsigma[ipix]).pdf(r) / pixarea
>>> dp_dV
3.1173200109121657e-05
\end{lstlisting}

\subsection{Marginal Distance Distribution Integrated over the Sky}

As our next example, we compute the marginal distance distribution, the probability density per unit distance integrated over the entire sky:
\begin{lstlisting}
>>> r = np.linspace(0, 150)
>>> dp_dr = [np.sum(prob * rr**2 * distnorm
... * norm(distmu, distsigma).pdf(rr)) for rr in r]
\end{lstlisting}

Finally, we plot the result:
\begin{lstlisting}
>>> plt.plot(r, dp_dr)
>>> plt.xlabel('distance (Mpc)')
>>> plt.ylabel('prob Mpc$^{-1}$')
>>> plt.show()
\end{lstlisting}
\includegraphics[width=\columnwidth]{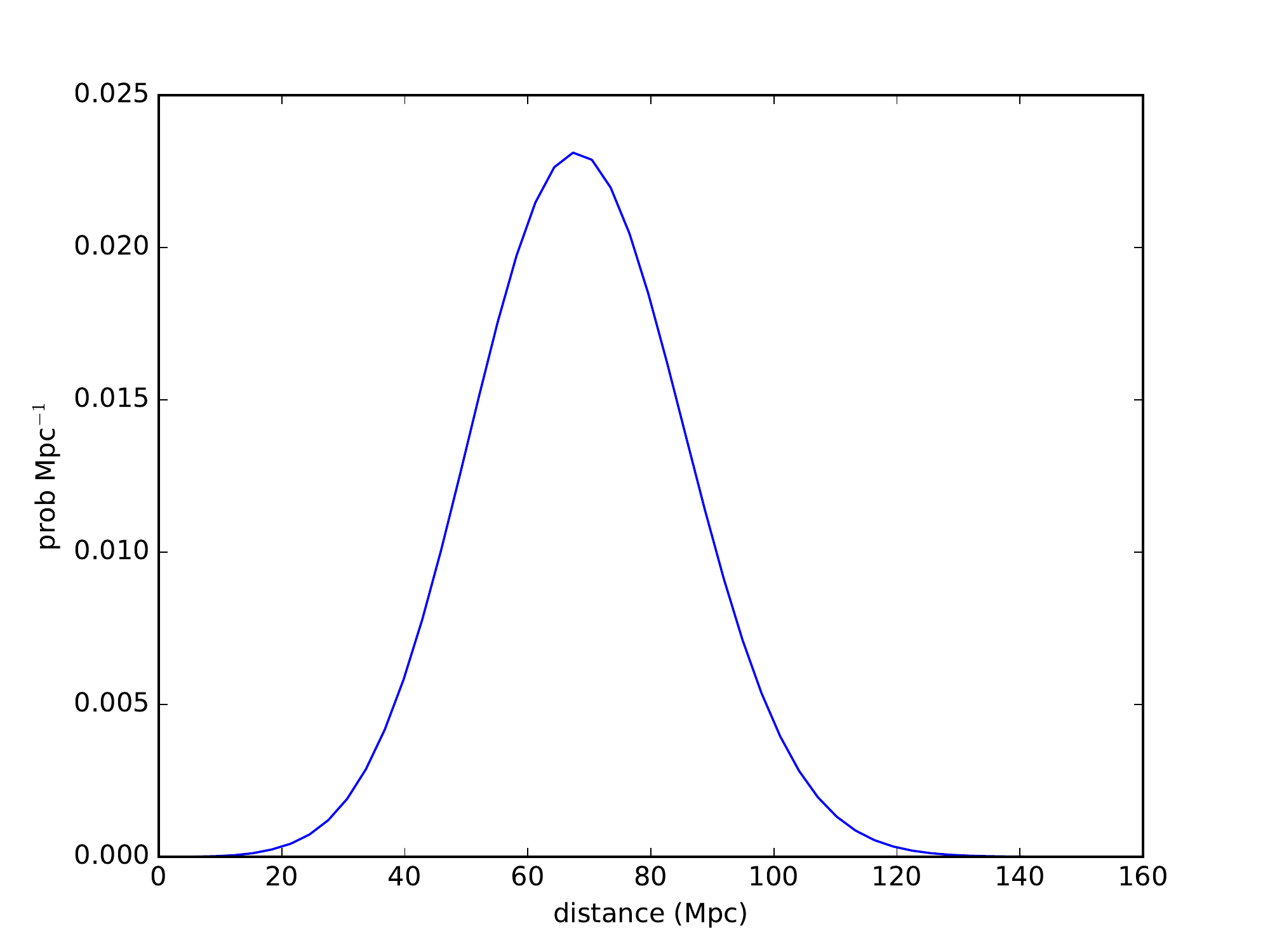}

\subsection{Ranked List of Galaxies}

As our final example, we will generate a ranked list of galaxies.

For the purpose of this demonstration, we will use the \acl{2MRS} (\acsu{2MRS}; \citealt{2MRS}) because it is a flux-limited all-sky spectroscopic redshift catalog. This greatly simplifies the issues of completeness, sky coverage, and accuracy of redshift estimates. First, download the entire catalog from VizieR \citep{vizier} using Astroquery:
\begin{lstlisting}
>>> from astroquery.vizier import Vizier
>>> Vizier.ROW_LIMIT = -1
>>> cat, = Vizier.get_catalogs('J/ApJS/199/26/table3')
\end{lstlisting}

According to \citet{2MASSGalaxyGroups}, the \ac{2MRS} luminosity function is well fit by a Schechter function with a cutoff absolute magnitude of $M_K^* = -23.55$ and a power-law index of $\alpha_K = -1$. We find the maximum absolute magnitude $M_K^\mathrm{max}$ for a completeness fraction of 0.5:
\begin{lstlisting}
>>> from scipy.special import gammaincinv
>>> completeness = 0.5
>>> alpha = -1.0
>>> MK_star = -23.55
>>> MK_max = MK_star + 2.5 * np.log10(
... gammaincinv(alpha + 2, completeness))
>>> MK_max
-23.947936347387156
\end{lstlisting}

We select only galaxies with positive redshifts and absolute magnitudes greater than $M_K^\mathrm{max}$:
\begin{lstlisting}
>>> from astropy.cosmology import WMAP9 as cosmo
>>> from astropy.table import Column
>>> import astropy.units as u
>>> import astropy.constants as c
>>> z = (u.Quantity(cat['cz']) / c.c).to(
... u.dimensionless_unscaled)
>>> MK = cat['Ktmag'] - cosmo.distmod(z)
>>> keep = (z > 0) & (MK < MK_max)
>>> cat = cat[keep]
>>> z = z[keep]
\end{lstlisting}

Then, we calculate the luminosity distance and \ac{HEALPix} index of each galaxy:
\begin{lstlisting}
>>> r = cosmo.luminosity_distance(z).to('Mpc').value
>>> theta = 0.5 * np.pi - cat['_DEJ2000'].to('rad').value
>>> phi = cat['_RAJ2000'].to('rad').value
>>> ipix = hp.ang2pix(nside, theta, phi)
\end{lstlisting}

We find the probability density per unit volume at the position of each galaxy:
\begin{lstlisting}
>>> dp_dV = prob[ipix] * distnorm[ipix] * norm(
... distmu[ipix], distsigma[ipix]).pdf(r) / pixarea
\end{lstlisting}

Finally, we sort the galaxies by descending probability density and take the top 50:
\begin{lstlisting}
>>> top50 = cat[np.flipud(np.argsort(dp_dV))][:50]
>>> top50['_RAJ2000', '_DEJ2000', 'Ktmag']
<Table masked=True length=50>
 _RAJ2000  _DEJ2000  Ktmag 
   deg       deg      mag  
 float64   float64  float32
--------- --------- -------
344.01190  36.36136   8.772
343.81122  36.67177   9.958
137.19089 -38.60788   9.566
334.86545  29.39581   9.835
359.81589  46.88923   9.307
  0.00695  47.27456   9.499
      ...       ...     ...
123.16494 -16.05073   9.727
341.26642  33.99616   9.799
339.33075  34.44790   9.204
137.27219 -35.90259  10.822
188.13953 -68.53886   9.609
339.01483  33.97575  10.032
\end{lstlisting}

\section{Volume Reconstruction Algorithm}
\label{sec:algorithm}

The volume reconstruction algorithm consists of a computationally trivial postprocessing stage that is added to the two established LIGO/Virgo methods for localization of \ac{CBC} events, the \acs{BAYESTAR} rapid triangulation code \citep{FirstTwoYears,leo-singer-thesis}, and the LALInference parameter estimation pipeline \citep{S6PE}.

The conditional mean and standard deviation of distance are extracted from \ac{BAYESTAR} as described in \S\ref{sec:bayestar} and from LALInference as explained in \S\ref{sec:lalinference} below. Then, the mean and standard deviation are converted to the ansatz parameters as described in \S\ref{sec:method-of-moments}.

\subsection{Volume Reconstruction in BAYESTAR}
\label{sec:bayestar}

\ac{BAYESTAR} \citep{FirstTwoYears,leo-singer-thesis} is a rapid position reconstruction algorithm for \ac{BNS} mergers. Its inputs are a trio of numbers for each detector: the matched\nobreakdashes-filter estimates of the arrival time, phase, and amplitude at each \ac{GW} site. It marginalizes over polarization angle, inclination angle, coalescence time, and distance by performing low-order Gaussian quadrature integration in a series of nested loops. The output is a \acs{HEALPix} all\nobreakdashes-sky map of posterior probability, consisting of $N_\mathrm{pix}$ equal\nobreakdashes-area pixels.

The \acs{BAYESTAR} distance prior is a power law of $r^k$, with $k$ being supplied by the user and normally set to $k=2$ for a spatially homogeneous source population. To evaluate the distances, we run \acs{BAYESTAR} two more times, with $k^\prime = k + 1$ and $k^{\prime\prime} = k + 2$. The resulting three sky maps are denoted $\langle 1 \rangle$, $\langle r \rangle$, and $\langle r^2 \rangle$. The HEALPix-sampled marginal sky posterior $\hat \rho$, conditional mean distance $\hat m$, and conditional standard deviation of distance $\hat s$ are then given by
\begin{align}
    \label{eq:bayestar-rho}    \hat\rho &= \langle 1 \rangle, \\
    \label{eq:bayestar-m}    \hat m &= \langle r \rangle / \langle 1 \rangle, \text{ and} \\
    \label{eq:bayestar-s}    \hat s &= \sqrt{\langle r^2 \rangle / \langle 1 \rangle - \hat m^2}.
\end{align}

Finally, the moments $\hat m$ and $\hat s$ are converted to the ansatz parameters $\hat\mu$, $\hat\sigma$, and $\hat N$ using the procedure described in \S\ref{sec:method-of-moments} below.

\acs{BAYESTAR} takes about a minute to run \citep{leo-singer-thesis}; the conventional one\nobreakdashes-dimensional sky map is ready with a response time of a few minutes. Since we will now run \acs{BAYESTAR} three times, the total number of computations will increase by about a factor of 3. Fortunately, since \acs{BAYESTAR} is able to make effective use of many CPU cores, we can offset the modest increase in computational cost by moving the analysis to a machine with more cores, resulting in a negligible overall change in running time.

\subsection{Volume Reconstruction in LALInference}
\label{sec:lalinference}

LALInference \citep{S6PE} is the Advanced \acs{LIGO} Bayesian parameter estimation library. It includes several algorithms that perform full modeling of the \ac{GW} signal and stochastic sampling of the \ac{CBC} parameter space. The inputs to LALInference are the \ac{GW} time series from all of the detectors. The output is a cloud of sample points drawn from the \ac{GW} posterior.

The samples are converted to a smooth multidimensional probability distribution by clustering them into $N$ disjoint sets, each consisting of $N_i$ spatially neighboring points, and building a \ac{KDE} for each cluster.\footnote{\url{https://github.com/farr/skyarea}} In Cartesian coordinates, the smoothed distribution is given by a double sum over the clusters and the samples within each cluster:
\begin{multline}
    \label{eq:kde}
    p(\boldsymbol x) = \sum_{i\,=\,1}^{N} W_i \left|2 \pi \boldsymbol C_i\right|^{-1/2}
    {N_i}^{-1} \\
    \sum_{j\,=\,1}^{N_i} \exp \left[ -\frac{1}{2}
    \left(\boldsymbol x - \boldsymbol X_{ij}\right)\transpose
    {{\boldsymbol C}_i}^{-1}
    \left(\boldsymbol x - \boldsymbol X_{ij}\right)
    \right].
\end{multline}
Here, $W_i$ is a weight associated with cluster $i$, and each cluster is described by its \ac{KDE} covariance ${\boldsymbol C}_i$ and $N_i$ samples $\boldsymbol X_{ij}$. The $\left|\ldots\right|$ denotes the matrix determinant.

The stochastic sampling takes hours to weeks depending on the sophistication of the waveform models that are used and on the treatment of uncertainty in detector calibration. It takes up to tens of minutes to build the \ac{KDE}.

We can exactly calculate the conditional mean and standard deviation of the distance for the \ac{KDE} posterior. First, we evaluate Eq.~(\ref{eq:kde}) at the position $\boldsymbol{x} = r \boldsymbol{n}$:
\begin{align}
    p(r \boldsymbol{n}) &= \sum_{i\,=\,1}^{N} \left(2 \pi c_i\right)^{-1/2}
    {N_i}^{-1} \sum_{j\,=\,1}^{N_i} w_{ij}
    \exp \left[ \frac{-(r - x_{ij})^2}{2 c_i} \right],
\intertext{with}
    c_i &= (\boldsymbol{n} \transpose {\boldsymbol C_i}^{-1} \boldsymbol{n})^{-1}, \\
    x_{ij} &= (\boldsymbol{n} \transpose {\boldsymbol C_i}^{-1} \boldsymbol X_{ij}) c_i, \text{ and} \\
    w_{ij} &= \frac{1}{2\pi} \sqrt{\frac{c_i}{\left| \boldsymbol C_i\right|}} \exp\left[\frac{1}{2}\left(\frac{{x_{ij}}^2}{c_i} - \boldsymbol X_{ij}\transpose \boldsymbol{C_i}^{-1}\boldsymbol X_{ij}\right)\right] W_i.
\end{align}
We compute the integrals of 1, $r$, and $r^2$, weighted by $p(r \boldsymbol{n}) r^2$:
\begin{align}
   \label{eq:kde-1}   \langle 1 \rangle &= \int_0^\infty p(r \boldsymbol{n}) r^2 \,dr = \sum_{ij} w_{ij} \langle{1}_{ij}\rangle / N_i, \\
   \label{eq:kde-r}   \langle r \rangle &= \int_0^\infty p(r \boldsymbol{n}) r^3 \,dr = \sum_{ij} w_{ij} \langle{r}_{ij}\rangle / N_i, \\
   \label{eq:kde-r2}   \langle r^2 \rangle &= \int_0^\infty p(r \boldsymbol{n}) r^4 \,dr = \sum_{ij} w_{ij} \langle{r^2}\rangle_{ij} / N_i,
\intertext{with}
   \langle{1}\rangle_{ij} &= \left({x_{ij}}^2 + c_i\right) a + x_{ij} b, \\
   \langle{r}\rangle_{ij} &= \left({x_{ij}}^3 + 3 x_{ij} c_i\right) a + \left({x_{ij}}^2 + 2 c_i\right) b, \\
   \langle{r^2}\rangle_{ij} &= \left({x_{ij}}^4 + 6 {x_{ij}}^2 c_i + 3 {c_i}^2\right) a + \left({x_{ij}}^3 + 5 x_{ij} c_i\right) b, \\
   a &= \frac{1}{2}\left(1 + \erf\left[\frac{x_{ij}}{\sqrt{2c_i}}\right]\right), \text{ and} \\
   b &= \sqrt{\frac{c_i}{2\pi}} \exp\left[\frac{-{x_{ij}}^2}{2c_i}\right].
\end{align}

Then $\langle 1 \rangle$, $\langle r \rangle$, and $\langle r^2 \rangle$ are converted to $\hat\rho$, $\hat m$, and $\hat s$ using Equations~(\ref{eq:bayestar-rho})\nobreakdashes--(\ref{eq:bayestar-s}). Finally, $\hat{m}$ and $\hat{s}$ are converted to $\hat\mu$, $\hat\sigma$, and $\hat{N}$ using the procedure described in \S\ref{sec:method-of-moments} below.

\subsection{Method of Moments}
\label{sec:method-of-moments}

For both \acs{BAYESTAR} and LALInference, the parameters of the ansatz distribution are extracted using the method of moments. The ansatz is that the conditional distribution of distance is described by the function
\begin{align}
    \label{eq:conditional-distance-distribution-ansatz}
    p(r | \boldsymbol{n}) &= \frac{N(\boldsymbol{n})}{\sqrt{2\pi}\sigma{(\boldsymbol{n})}} \exp\left[-\frac{\left(r - \mu(\boldsymbol{n})\right)^2}{2\sigma(\boldsymbol{n})^2}\right] r^2 \\
    \text{for } r &\geq 0. \nonumber
\end{align}
The $n$th moment of the distance ansatz is
\begin{equation}
    \overline{r^n}(\mu, \sigma) = \int_0^\infty \frac{N}{\sqrt{2\pi}\sigma} \exp\left[-\frac{\left(r - \mu\right)^2}{2\sigma^2}\right] r^{2+n} \, \dd r.
\end{equation}
The conditional mean and standard deviation of the ansatz distribution are
\begin{align}
    m(\mu, \sigma) &= \overline{r}(\mu, \sigma) \text{ and} \\
    s(\mu, \sigma) &= \sqrt{\overline{r^2}(\mu, \sigma) - \overline{r}^2(\mu, \sigma)}.
\end{align}
Our task is, given the conditional mean $\hat{m}$ and standard deviation $\hat{s}$ as measured from the actual posterior probability distribution, to numerically solve the following system of equations for $\hat\mu$ and $\hat\sigma$:
\begin{align}
    \label{eq:hat-m}
    \hat{m} &= \overline{r}(\hat\mu, \hat\sigma) \text{ and} \\
    \label{eq:hat-s}
    \hat{s} &= \sqrt{\overline{r^2}(\hat\mu, \hat\sigma) - \overline{r}^2(\hat\mu, \hat\sigma)}.
\end{align}

We can reduce this to a single equation by defining $z = \mu / \sigma$ and $\hat{z} = \hat\mu / \hat\sigma$. With this substitution, the moments can be written as
\begin{align*}
    \overline{r^n} &= N Q(-z) \sigma^{2 + n} x_{2 + n}(z), \\
\intertext{with}
    x_2(z) &= z^2 + 1 + z H(-z), \\
    x_3(z) &= z^3 + 3z + (z^2 + 2) H(-z), \text{ and} \\
    x_4(z) &= z^4 + 6z^2 + 3 + (z^3 + 5z) H(-z).
\end{align*}
The function $Q(x) = \erfc(x/\sqrt{2})/2$ is the upper tail of the normal distribution, $P(x)=\exp\left(-x^2/2\right)/\sqrt{2\pi}$ is the normal distribution function, and $H(x) = P(x)/Q(x)$ is the hazard function.
Then Equations~(\ref{eq:hat-m}) and (\ref{eq:hat-s}) become
\begin{equation}
    \label{eq:mom-f}
    f(\hat{z}) = \left(1 + \left(\frac{\hat{s}}{\hat{m}}\right)^2\right)
        {x_3(\hat{z})}^2 - x_2(\hat{z}) x_4(\hat{z}) = 0.
\end{equation}

The derivative of the left-hand side, $f^\prime(\hat{z})$, is given by
\begin{multline}
    \label{eq:mom-fprime}
    f^\prime(\hat{z}) = 2 \left(1 + \left(\frac{\hat{s}}{\hat{m}}\right)^2\right)
        x_3(\hat{z}) {x_3}^\prime(\hat{z}) \\
        - x_2(\hat{z}) {x_4}^\prime(\hat{z}) - {x_2}^\prime(\hat{z}) x_4(\hat{z})
\end{multline}
with
\begin{align*}
    x_2^\prime(z) &= 2z + H(-z) + z \partial_z H(-z), \\
    x_3^\prime(z) &= 3z^2 + 3 + 2 z H(-z) + (z^2 + 2), \partial_z H(-z) \\
    x_4^\prime(z) &= 4z^3 + 12z + (3z^2 + 5) H(-z) + (z^3 + 5z) \partial_z H(-z), \\
    \text{and } &\partial_z H(-z) = -H(-z) (z + H(-z)).
\end{align*}

We solve Equations~(\ref{eq:mom-f}) and (\ref{eq:mom-fprime}) for $\hat{z}$ using Steffensen's method\footnote{The  \texttt{gsl\_root\_fdfsolver\_steffenson} method in the GNU Scientific Library.}, an accelerated Newton solver. Starting from an initial value of $\hat{z}_0 = \hat{m}/\hat{s}$, the solution converges to machine precision in 10 iterations.

Finally, we calculate $\hat\mu$, $\hat\sigma$, and $N$ as follows:
\begin{align*}
    \hat\sigma &= m x_2(\hat{z}) / x_3(\hat{z}), \\
    \hat\mu &= \hat\sigma \hat{z}, \\
    \hat{N} &= \left(Q(-\hat{z}) \hat\sigma^2 x_2(\hat{z})\right)^{-1}.
\end{align*}
In the rare event that the solution does not converge, or yields an invalid value such that $\overline{r^2}(\hat\mu, \hat\sigma) - \overline{r}^2(\hat\mu, \hat\sigma) < 0$, we set
\begin{align*}
    \hat\mu &= \infty, \\
    \hat\sigma &= 1, \\
    \hat{N} &= 0.
\end{align*}

\section{Faithfulness}
\label{sec:faithfulness}

The ansatz guarantees that the first two moments of distance are exactly reproduced along all \acp{LOS}. However, we must ask how accurately the ansatz represents the 3D posterior as a whole. The $P$--$P$ plot graphical test, popularized in the \ac{GW} parameter estimation literature by \citet{SiderySkyLocalizationComparison}, compares two populations by plotting their cumulative distributions against each other. If the two distributions match, then the result should be a diagonal line.

\begin{figure*}
    \begin{minipage}{0.5\textwidth}
        \begin{center}
            \includegraphics[width=\textwidth]{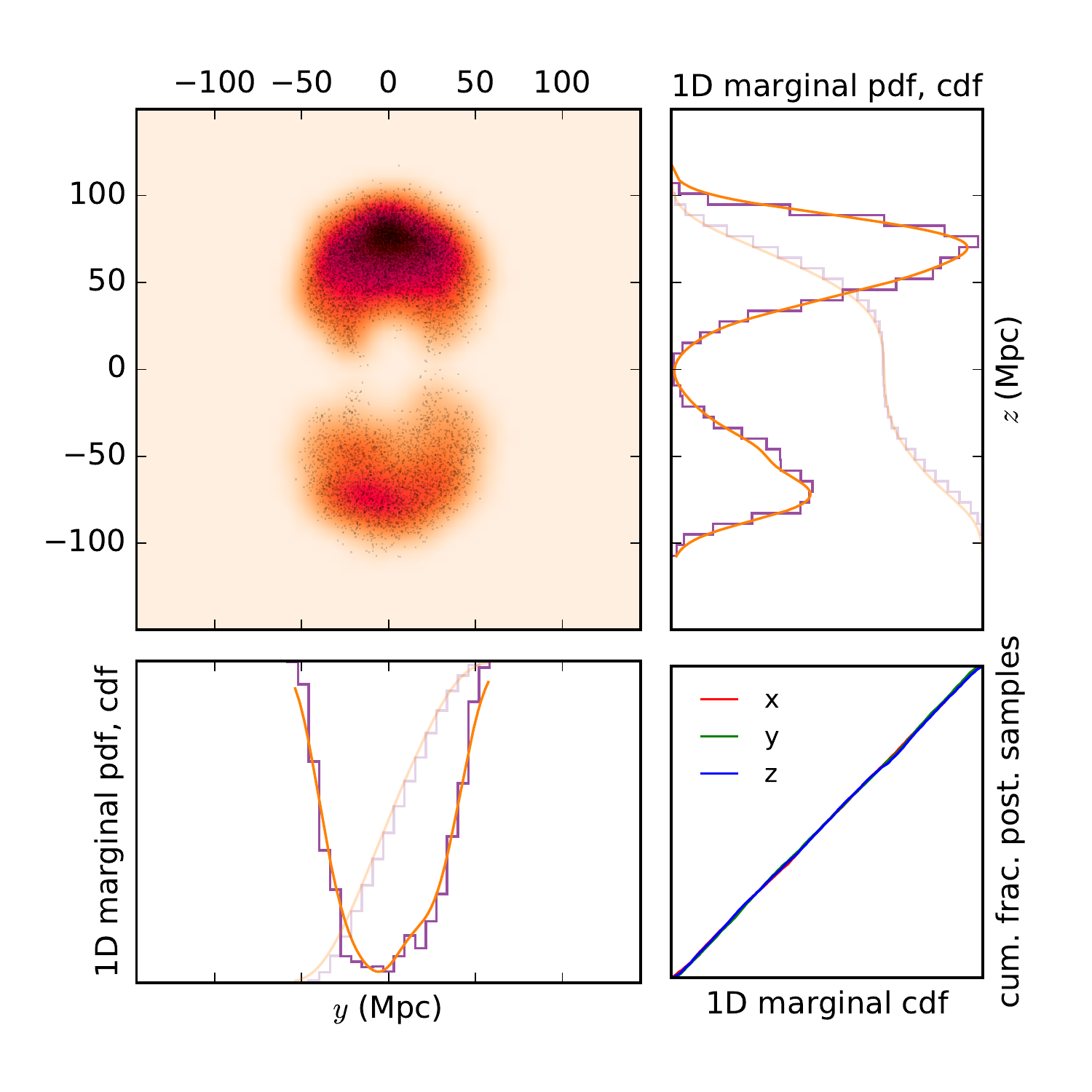} \\
            (a) \ac{KDE} versus posterior samples
        \end{center}
    \end{minipage}
    \begin{minipage}{0.5\textwidth}
        \begin{center}
            \includegraphics[width=\textwidth]{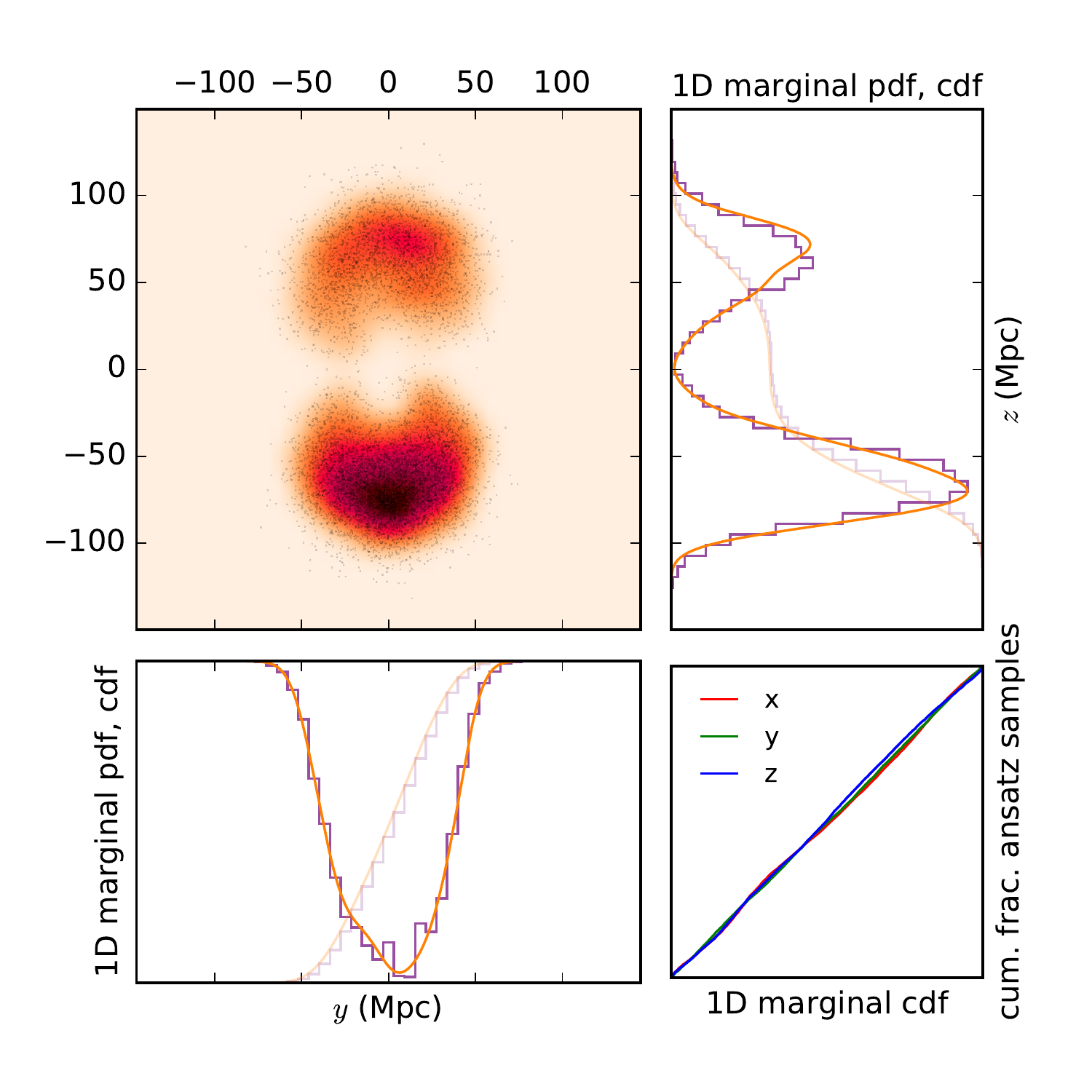} \\
            (b) \ac{KDE} versus samples from ansatz
        \end{center}
    \end{minipage}
    \caption{\label{fig:kde-ansatz-comparison}Comparison between the posterior sample chain, the \ac{KDE}, and the ansatz distribution. In panel~(a), the heat map shows a 2D projection of the \ac{KDE}. The black dots are the LALInference posterior samples from which the \ac{KDE} was built. The top right and bottom left plots show the 1D projections. The dark, smooth, orange line is the 1D marginal \ac{KDE} and the dark, purple, stepped line is a 1D histogram of the LALInference posterior samples. The faint lines of the corresponding colors and styles are the respective 1D cumulative distributions. The bottom right plot is a $P$--$P$ plot of the 1D cumulative distribution of the \ac{KDE} versus the 1D cumulative histogram of the posterior samples. Panel~(b) is the same as panel~(a), except that samples from the ansatz distribution are substituted for samples from the posterior.}
\end{figure*}

In our case, we compare the \ac{KDE} to the LALInference posterior samples by projecting both the \ac{KDE} and the posterior samples along the distribution's three principal axes, yielding three $P$--$P$ tests. As shown in Fig.~\ref{fig:kde-ansatz-comparison}a, the plot is nearly diagonal, indicating that the \ac{KDE} is a faithful representation of the posterior samples. We then compare the \ac{KDE} with the ansatz by drawing samples from the ansatz distribution (Fig.~\ref{fig:kde-ansatz-comparison}b). Some deviation is perceptible; in the most extreme cases we find a maximum difference in credible levels of about 5\%. $P$--$P$ tests of the conditional distance distribution itself along individual \acp{LOS} generally also agree within 5\% or better, except in directions of low probability (small $\rho_i$).

\begin{figure*}
    \begin{minipage}{0.5\textwidth}
        \begin{center}
            \includegraphics[width=\textwidth]{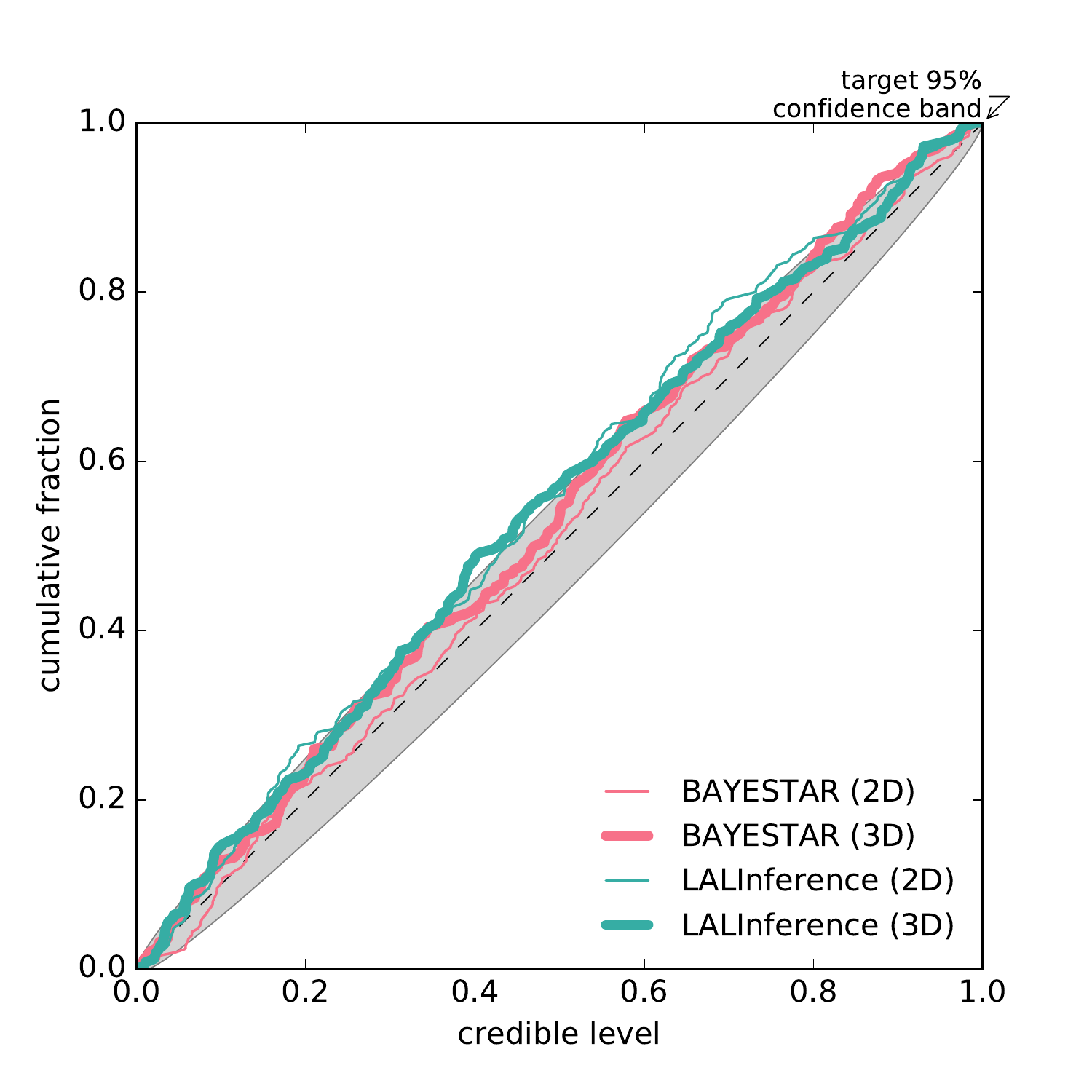} \\
            (a) Ensemble $P$--$P$ plot (\ac{O1})
        \end{center}
    \end{minipage}
    \begin{minipage}{0.5\textwidth}
        \begin{center}
            \includegraphics[width=\textwidth]{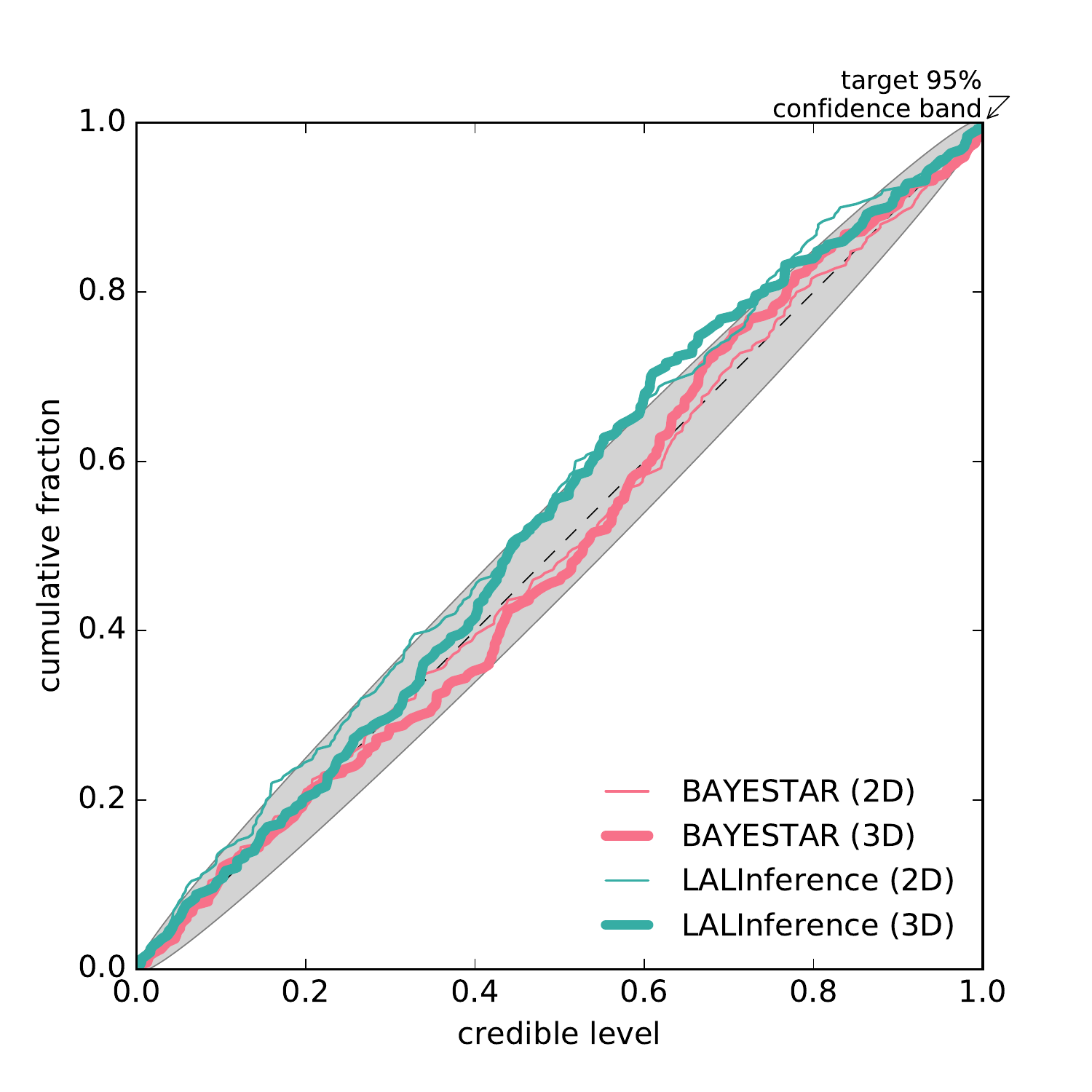} \\
            (b) Ensemble $P$--$P$ plot (\ac{O2})
        \end{center}
    \end{minipage}
    \caption{\label{fig:pp}Ensemble $P$--$P$ plot test for the 250 events from the \ac{O1} scenario~(a) and the 250 events from the \ac{O2} scenario~(b).}
\end{figure*}

We test the statistical self-consistency of the entire ensemble of simulated events in Fig.~\ref{fig:pp}. Here, we show a cumulative histogram of the number of simulated events whose true \ac{2D} and \ac{3D} coordinates are found within a given credible level. We find that both the \ac{2D} sky maps and the \ac{3D} ansatz are self-consistent within a binomial 95\% tolerance band due to the finite sample size of 250 events.

Our interpretation is that the ansatz is a reasonable approximation of the full 3D posterior, in the sense that a stated 50\% credible volume has a 50\%$\pm 5$\% chance of containing the source. The most obvious alternative to the ansatz is a densely sampled 3D grid, or a stack of 2D sky maps for a series of distance shells. Either would be just as conceptually simple, but computationally cumbersome due to size. The \ac{KDE} is an accurate representation of the posterior, but is expensive to evaluate because it is a sum of $10^4-10^5$ Gaussians. As a rapidly available data product and as a tool for real\nobreakdashes-time observation planning, the ansatz distribution is a reasonable compromise.

\providecommand{\acrolowercase}[1]{\lowercase{#1}}

\begin{acronym}
\acro{2D}[2D]{two\nobreakdashes-dimensional}
\acro{2+1D}[2+1D]{2+1\nobreakdashes--dimensional}
\acro{2MRS}[2MRS]{2MASS Redshift Survey}
\acro{3D}[3D]{three\nobreakdashes-dimensional}
\acro{2MASS}[2MASS]{Two Micron All Sky Survey}
\acro{AdVirgo}[AdVirgo]{Advanced Virgo}
\acro{AMI}[AMI]{Arcminute Microkelvin Imager}
\acro{AGN}[AGN]{active galactic nucleus}
\acroplural{AGN}[AGN\acrolowercase{s}]{active galactic nuclei}
\acro{aLIGO}[aLIGO]{Advanced \acs{LIGO}}
\acro{ASKAP}[ASKAP]{Australian \acl{SKA} Pathfinder}
\acro{ATCA}[ATCA]{Australia Telescope Compact Array}
\acro{ATLAS}[ATLAS]{Asteroid Terrestrial-impact Last Alert System}
\acro{BAT}[BAT]{Burst Alert Telescope\acroextra{ (instrument on \emph{Swift})}}
\acro{BATSE}[BATSE]{Burst and Transient Source Experiment\acroextra{ (instrument on \acs{CGRO})}}
\acro{BAYESTAR}[BAYESTAR]{BAYESian TriAngulation and Rapid localization}
\acro{BBH}[BBH]{binary black hole}
\acro{BHBH}[BHBH]{\acl{BH}\nobreakdashes--\acl{BH}}
\acro{BH}[BH]{black hole}
\acro{BNS}[BNS]{binary neutron star}
\acro{CARMA}[CARMA]{Combined Array for Research in Millimeter\nobreakdashes-wave Astronomy}
\acro{CASA}[CASA]{Common Astronomy Software Applications}
\acro{CFH12k}[CFH12k]{Canada--France--Hawaii $12\,288 \times 8\,192$ pixel CCD mosaic\acroextra{ (instrument formerly on the Canada--France--Hawaii Telescope, now on the \ac{P48})}}
\acro{CRTS}[CRTS]{Catalina Real-time Transient Survey}
\acro{CTIO}[CTIO]{Cerro Tololo Inter-American Observatory}
\acro{CBC}[CBC]{compact binary coalescence}
\acro{CCD}[CCD]{charge coupled device}
\acro{CDF}[CDF]{cumulative distribution function}
\acro{CGRO}[CGRO]{Compton Gamma Ray Observatory}
\acro{CMB}[CMB]{cosmic microwave background}
\acro{CRLB}[CRLB]{Cram\'{e}r\nobreakdashes--Rao lower bound}
\acro{cWB}[\acrolowercase{c}WB]{Coherent WaveBurst}
\acro{DASWG}[DASWG]{Data Analysis Software Working Group}
\acro{DBSP}[DBSP]{Double Spectrograph\acroextra{ (instrument on \acs{P200})}}
\acro{DCT}[DCT]{Discovery Channel Telescope}
\acro{DECAM}[DECam]{Dark Energy Camera\acroextra{ (instrument on the Blanco 4\nobreakdashes-m telescope at \acs{CTIO})}}
\acro{DES}[DES]{Dark Energy Survey}
\acro{DFT}[DFT]{discrete Fourier transform}
\acro{EM}[EM]{electromagnetic}
\acro{ER8}[ER8]{eighth engineering run}
\acro{FD}[FD]{frequency domain}
\acro{FAR}[FAR]{false alarm rate}
\acro{FFT}[FFT]{fast Fourier transform}
\acro{FIR}[FIR]{finite impulse response}
\acro{FITS}[FITS]{Flexible Image Transport System}
\acro{FLOPS}[FLOPS]{floating point operations per second}
\acro{FOV}[FOV]{field of view}
\acroplural{FOV}[FOV\acrolowercase{s}]{fields of view}
\acro{FTN}[FTN]{Faulkes Telescope North}
\acro{FWHM}[FWHM]{full width at half-maximum}
\acro{GBM}[GBM]{Gamma-ray Burst Monitor\acroextra{ (instrument on \emph{Fermi})}}
\acro{GCN}[GCN]{Gamma-ray Coordinates Network}
\acro{GMOS}[GMOS]{Gemini Multi-object Spectrograph\acroextra{ (instrument on the Gemini telescopes)}}
\acro{GRB}[GRB]{gamma-ray burst}
\acro{GSC}[GSC]{Gas Slit Camera}
\acro{GSL}[GSL]{GNU Scientific Library}
\acro{GTC}[GTC]{Gran Telescopio Canarias}
\acro{GW}[GW]{gravitational wave}
\acro{HAWC}[HAWC]{High\nobreakdashes-Altitude Water \v{C}erenkov Gamma\nobreakdashes-Ray Observatory}
\acro{HCT}[HCT]{Himalayan Chandra Telescope}
\acro{HEALPix}[HEALP\acrolowercase{ix}]{Hierarchical Equal Area isoLatitude Pixelization}
\acro{HEASARC}[HEASARC]{High Energy Astrophysics Science Archive Research Center}
\acro{HETE}[HETE]{High Energy Transient Explorer}
\acro{HFOSC}[HFOSC]{Himalaya Faint Object Spectrograph and Camera\acroextra{ (instrument on \acs{HCT})}}
\acro{HMXB}[HMXB]{high\nobreakdashes-mass X\nobreakdashes-ray binary}
\acroplural{HMXB}[HMXB\acrolowercase{s}]{high\nobreakdashes-mass X\nobreakdashes-ray binaries}
\acro{HSC}[HSC]{Hyper Suprime\nobreakdashes-Cam\acroextra{ (instrument on the 8.2\nobreakdashes-m Subaru telescope)}}
\acro{IACT}[IACT]{imaging atmospheric \v{C}erenkov telescope}
\acro{IIR}[IIR]{infinite impulse response}
\acro{IMACS}[IMACS]{Inamori-Magellan Areal Camera \& Spectrograph\acroextra{ (instrument on the Magellan Baade telescope)}}
\acro{IMR}[IMR]{inspiral-merger-ringdown}
\acro{IPAC}[IPAC]{Infrared Processing and Analysis Center}
\acro{IPN}[IPN]{InterPlanetary Network}
\acro{IPTF}[\acrolowercase{i}PTF]{intermediate \acl{PTF}}
\acro{ISM}[ISM]{interstellar medium}
\acro{ISS}[ISS]{International Space Station}
\acro{KAGRA}[KAGRA]{KAmioka GRAvitational\nobreakdashes-wave observatory}
\acro{KDE}[KDE]{kernel density estimator}
\acro{KN}[KN]{kilonova}
\acroplural{KN}[KNe]{kilonovae}
\acro{LAT}[LAT]{Large Area Telescope}
\acro{LCOGT}[LCOGT]{Las Cumbres Observatory Global Telescope}
\acro{LHO}[LHO]{\ac{LIGO} Hanford Observatory}
\acro{LIB}[LIB]{LALInference Burst}
\acro{LIGO}[LIGO]{Laser Interferometer \acs{GW} Observatory}
\acro{llGRB}[\acrolowercase{ll}GRB]{low\nobreakdashes-luminosity \ac{GRB}}
\acro{LLOID}[LLOID]{Low Latency Online Inspiral Detection}
\acro{LLO}[LLO]{\ac{LIGO} Livingston Observatory}
\acro{LMI}[LMI]{Large Monolithic Imager\acroextra{ (instrument on \ac{DCT})}}
\acro{LOFAR}[LOFAR]{Low Frequency Array}
\acro{LOS}[LOS]{line of sight}
\acroplural{LOS}[LOSs]{lines of sight}
\acro{LMC}[LMC]{Large Magellanic Cloud}
\acro{LSB}[LSB]{long, soft burst}
\acro{LSC}[LSC]{\acs{LIGO} Scientific Collaboration}
\acro{LSO}[LSO]{last stable orbit}
\acro{LSST}[LSST]{Large Synoptic Survey Telescope}
\acro{LT}[LT]{Liverpool Telescope}
\acro{LTI}[LTI]{linear time invariant}
\acro{MAP}[MAP]{maximum a posteriori}
\acro{MBTA}[MBTA]{Multi-Band Template Analysis}
\acro{MCMC}[MCMC]{Markov chain Monte Carlo}
\acro{MLE}[MLE]{\ac{ML} estimator}
\acro{ML}[ML]{maximum likelihood}
\acro{MOU}[MOU]{memorandum of understanding}
\acroplural{MOU}[MOUs]{memoranda of understanding}
\acro{MWA}[MWA]{Murchison Widefield Array}
\acro{NED}[NED]{NASA/IPAC Extragalactic Database}
\acro{NSBH}[NSBH]{neutron star\nobreakdashes--black hole}
\acro{NSBH}[NSBH]{\acl{NS}\nobreakdashes--\acl{BH}}
\acro{NSF}[NSF]{National Science Foundation}
\acro{NSNS}[NSNS]{\acl{NS}\nobreakdashes--\acl{NS}}
\acro{NS}[NS]{neutron star}
\acro{O1}[O1]{\acl{aLIGO}'s first observing run}
\acro{O2}[O2]{\acl{aLIGO}'s second observing run}
\acro{oLIB}[\acrolowercase{o}LIB]{Omicron+\acl{LIB}}
\acro{OT}[OT]{optical transient}
\acro{P48}[P48]{Palomar 48~inch Oschin telescope}
\acro{P60}[P60]{robotic Palomar 60~inch telescope}
\acro{P200}[P200]{Palomar 200~inch Hale telescope}
\acro{PC}[PC]{photon counting}
\acro{PESSTO}[PESSTO]{Public ESO Spectroscopic Survey of Transient Objects}
\acro{PSD}[PSD]{power spectral density}
\acro{PSF}[PSF]{point-spread function}
\acro{PS1}[PS1]{Pan\nobreakdashes-STARRS~1}
\acro{PTF}[PTF]{Palomar Transient Factory}
\acro{QUEST}[QUEST]{Quasar Equatorial Survey Team}
\acro{RAPTOR}[RAPTOR]{Rapid Telescopes for Optical Response}
\acro{REU}[REU]{Research Experiences for Undergraduates}
\acro{RMS}[RMS]{root mean square}
\acro{ROTSE}[ROTSE]{Robotic Optical Transient Search}
\acro{S5}[S5]{\ac{LIGO}'s fifth science run}
\acro{S6}[S6]{\ac{LIGO}'s sixth science run}
\acro{SAA}[SAA]{South Atlantic Anomaly}
\acro{SHB}[SHB]{short, hard burst}
\acro{SHGRB}[SHGRB]{short, hard \acl{GRB}}
\acro{SKA}[SKA]{Square Kilometer Array}
\acro{SMT}[SMT]{Slewing Mirror Telescope\acroextra{ (instrument on \acs{UFFO} Pathfinder)}}
\acro{SNR}[S/N]{signal\nobreakdashes-to\nobreakdashes-noise ratio}
\acro{SSC}[SSC]{synchrotron self\nobreakdashes-Compton}
\acro{SDSS}[SDSS]{Sloan Digital Sky Survey}
\acro{SED}[SED]{spectral energy distribution}
\acro{SGRB}[SGRB]{short \acl{GRB}}
\acro{SN}[SN]{supernova}
\acroplural{SN}[SN\acrolowercase{e}]{supernova}
\acro{SNIa}[\acs{SN}\,I\acrolowercase{a}]{Type~Ia \ac{SN}}
\acroplural{SNIa}[\acsp{SN}\,I\acrolowercase{a}]{Type~Ic \acp{SN}}
\acro{SNIcBL}[\acs{SN}\,I\acrolowercase{c}\nobreakdashes-BL]{broad\nobreakdashes-line Type~Ic \ac{SN}}
\acroplural{SNIcBL}[\acsp{SN}\,I\acrolowercase{c}\nobreakdashes-BL]{broad\nobreakdashes-line Type~Ic \acp{SN}}
\acro{SVD}[SVD]{singular value decomposition}
\acro{TAROT}[TAROT]{T\'{e}lescopes \`{a} Action Rapide pour les Objets Transitoires}
\acro{TDOA}[TDOA]{time delay on arrival}
\acroplural{TDOA}[TDOA\acrolowercase{s}]{time delays on arrival}
\acro{TD}[TD]{time domain}
\acro{TOA}[TOA]{time of arrival}
\acroplural{TOA}[TOA\acrolowercase{s}]{times of arrival}
\acro{TOO}[TOO]{target\nobreakdashes-of\nobreakdashes-opportunity}
\acroplural{TOO}[TOO\acrolowercase{s}]{targets of opportunity}
\acro{UFFO}[UFFO]{Ultra Fast Flash Observatory}
\acro{UHE}[UHE]{ultra high energy}
\acro{UVOT}[UVOT]{UV/Optical Telescope\acroextra{ (instrument on \emph{Swift})}}
\acro{VHE}[VHE]{very high energy}
\acro{VISTA}[VISTA@ESO]{Visible and Infrared Survey Telescope}
\acro{VLA}[VLA]{Karl G. Jansky Very Large Array}
\acro{VLT}[VLT]{Very Large Telescope}
\acro{VST}[VST@ESO]{\acs{VLT} Survey Telescope}
\acro{WAM}[WAM]{Wide\nobreakdashes-band All\nobreakdashes-sky Monitor\acroextra{ (instrument on \emph{Suzaku})}}
\acro{WCS}[WCS]{World Coordinate System}
\acro{WSS}[w.s.s.]{wide\nobreakdashes-sense stationary}
\acro{XRF}[XRF]{X\nobreakdashes-ray flash}
\acroplural{XRF}[XRF\acrolowercase{s}]{X\nobreakdashes-ray flashes}
\acro{XRT}[XRT]{X\nobreakdashes-ray Telescope\acroextra{ (instrument on \emph{Swift})}}
\acro{ZTF}[ZTF]{Zwicky Transient Facility}
\end{acronym}
 
\acknowledgements
We thank the Aspen Center for Physics and NSF grant \#1066293 for hospitality during the conception, writing, and editing of this paper. We thank P~Shawhan and F~Tombesi for detailed feedback on the manuscript. The online data release is available at \url{https://dcc.ligo.org/P1500071/public/html}. This is \acs{LIGO} document P1500071\nobreakdashes-v7.

\software{Astropy \citep{astropy}, GNU Scientific Library \citep{gsl}, HEALPix \citep{healpix}, Matplotlib \citep{matplotlib}, Yt \citep{yt}}

\bibliographystyle{aasjournal}

\end{document}